\documentclass[%
 reprint,
superscriptaddress,
 amsmath,amssymb,
 aps,
pra,
]{revtex4-2}

\usepackage{graphicx}% Include figure files
\usepackage{dcolumn}% Align table columns on decimal point
\usepackage{bm}% bold math
\usepackage{hyperref}% add hypertext capabilities
\usepackage[mathlines]{lineno}% Enable numbering of text and display math
\usepackage{epstopdf}
\usepackage{color}
\begin{document}

%\preprint{APS/123-QED}

\title{Electromagnetic fields between moving mirrors: \\ Singular waveforms inside Doppler cavities}

% Force line breaks with \\

\author{Theodoros T. Koutserimpas}
 \affiliation{Department of Electrical and Computer Engineering, \\
 Princeton University, Princeton NJ-08544, USA.}
 \email{tkoutserimpas@princeton.edu}
\author{Constantinos Valagiannopoulos}%
 \affiliation{Department of Physics, School of Sciences and Humanities\\ Nazarbayev University, Astana KZ-10000, Kazakhstan.}
 \email{valagiannopoulos@gmail.com}

\begin{abstract}
Phenomena of wave propagation in dynamically varying structures have reemerged as the temporal variations of the medium's properties can extend the possibilities for electromagnetic wave manipulation. While the dynamical change of the electromagnetic medium's properties is a difficult task, the movement of scatterers is not. In this paper, we analyze the electromagnetic fields trapped inside two smoothly moving mirrors. We employ the method of characteristics and take into account the relativistic phenomena to show that the temporally and spatially local Doppler effects can filter and amplify the electromagnetic signal, tailoring the $k-$ and $\omega-$content of the transients. It is shown using the Doppler factor  and the change of the distance between neighbor characteristics that the dynamical movement of the boundaries can lead to condensated characteristics resulting in field amplification or dilution of the characteristics resulting in the attenuation of the signal. In the case of periodically moving mirrors the field distribution is shown that asymptotically leads to exponentially growing delta-like wave packets at discrete points of space with a limiting number of peaks due to the fact that the velocity of the mechanical vibrations can not exceed that of light. The theoretical analysis is also verified by FDTD simulations and is connected with the theory of mode locking.
\end{abstract}

\maketitle

%\tableofcontents

\section{\label{sec:intro}Introduction}

The study of waves in dynamically changing boundaries is a physical problem that dates back to the early 20th century. Lord Rayleigh and Sir J. Larmor studied the dynamics of the radiation pressure \cite{rayleigh,larmor}. Their approach was utilized for the mathematical description of the motion of a string, with one end fixed and the other one moving \cite{havelock,nicolai}. Minkowski's space-time representation \cite{minkowski}, allowed for a geometrical approach to wave problems since the spatial and the temporal variables should be treated on equal footing.

Minkowski's space-time representation led to the method of characteristics which was first employed (to the authors' knowledge) for dynamically changing boundaries in \cite{balazs}, without taking into account the relativistic effects imparted by the velocity of the boundary. The case of incident electromagnetic waves in vibrating perfect conductors, accounting for relativistic effects was studied in \cite{cooper1,bladel,cooper2}. Furthermore in \cite{rafael1,rafael2}, the connection of the reflection times of the characteristics inside two boundaries was established with circle and torus maps providing additional mathematical intuition regarding the possible effects of the mirrors' periodic movement to the electromagnetic transients. Such studies induce a very limited physical intuition while attempt to describe a continuous phenomenon (wave propagation) as an identity that consists of only a limited number of points.

Although the objective of the presented work is restricted to classical electromagnetic theory and the Doppler effect, it is worth noticing that wave phenomena and systems with vibrating boundaries are strongly connected with the Casimir effect \cite{Dalvit1}. As suggested in \cite{Moore,Ji1,Dalvit2,Michael} and references therein, vacuum fluctuations can indeed lead to photon creation from the energy exchange of vacuum and the mechanical agent causing the boundary vibrations. 

In recent years, electromagnetic wave propagation in temporally dynamic media and systems has reemerged as a possible direction to enable interesting wave propagation properties and extend the possibilities to control electromagnetic waves \cite{Engheta1}. Intriguingly, the research in this field is mainly focused on the possible frequency changes, the energy growth and the nonreciprocal properties of waves under a temporal modulation of the medium's properties for bulk metamaterials or metasurfaces. For more details on the recent developments of electromagnetic waves in time-varying media consult the review article \cite{galiffi1} and its references therein.

A more pragmatic approach could be via the interaction of electromagnetic waves with moving obstacles. In such scenarios the Doppler effect dominates. As it is well known, if an object made of a perfect conductor moves towards an incident wave, the reflection undergoes a Doppler blue-shift and the reflection coefficient is greater than unity. If the object moves away from the incident wave the reflection undergoes a Doppler red-shift and the reflection coefficient is less than unity.

In this article, we study the interactions of electromagnetic transients inside two moving mirrors. We restrict our analysis to the classical electromagnetic phenomena and assume an existing square integrable profile of electromagnetic waves trapped inside two perfect electric conductors that start moving at a time $t=t_{0}$, with an arbitrary but smooth motion. Solutions are found using the method of characteristics and phenomena of energy growth or attenuation are interpreted by the change of the distance of neighbor characteristics as reflected by the moving boundaries. The special case of periodically moving boundaries is also examined, predicting the creation of ultrashort peaks which is verified by FDTD simulations (performed for the first time to the authors' knowledge for closed cavity systems) and conceptually connected with the theory of mode locking \cite{Siegman}.

More precisely, in Section (\ref{sec:theory}), the mathematical modeling of the electromagnetic transients, the electromagnetic energy and the Doppler factor are derived using the method of characteristics in the $(t,x)$ space. Phenomena of energy growth or attenuation and frequency shift are interpreted geometrically via the method of characteristics as the density of the characteristics condenses or dilutes at a point $(t,x)$ of reflection. In Section (\ref{subsec:periodic_boundaries_gen}), we solve for the special cases where the boundaries are moving periodically. Such cases have special properties when a group of characteristics is periodic in time. Such periodic characteristics, as thoroughly explained in the aforementioned Section, can be attractive and concentrate neighbor characteristics or repulsive and attenuate the neighbor transients. After a course of many periods these phenomena form an electromagnetic energy density of delta-like peaks, while the number of peaks is restricted by the maximum velocity of the vibrations, which can not exceed that of light. In Section (\ref{subsec:periodic_boundaries_cs1}), the specific solutions of the analysis of the characteristics is found for a sinusoidal movement of the boundaries. Section (\ref{sec:fdtd}), exhibits the chonophotographs of the electric field and the energy density distribution calculated by a 1D homemade Finite-Difference Time-Domain (FDTD) model and are in full agreement with the analysis of the previous Sections, while similarities with mode locking are described. In Section (\ref{sec:discussion}), we discuss the possible applications and implementations as well as the differences between the energy amplification mechanisms exhibited by the Doppler effect versus those found recently in time-periodic optical media.

\section{\label{sec:theory}The solution of characteristics}
\begin{figure}
\includegraphics{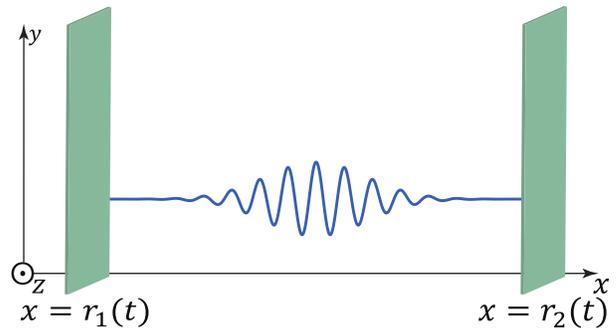}% Here is how to import EPS art
\caption{\label{fig:fig_1} Two perfect mirrors move perpendicular to the $x$ axis. Their position is a function of time: $x=r_{1}(t)$ and $x=r_{2}(t)$, respectively, while transient electromagnetic waves reflect back and forth.}
\end{figure}

Let us assume two parallel perfect mirrors, placed perpendicular on the $x$ axis. The two mirrors are assumed to be flat and totally reflective while constitute a Fabry-Perot cavity. 
 The mirrors are moving due to an external agent, with known displacement functions. The reference of observation is the stationary laboratory frame. The first mirror is at $x=r_{1}(t)$, while the second one is at $x=r_{2}(t)$ (as shown in Fig. \ref{fig:fig_1}). The space between the two plates is empty with ${\epsilon _0}$, ${\mu _0}$ and $c = {1 \mathord{\left/
 {\vphantom {1 {\sqrt {{\epsilon _0}{\mu _0}} }}} \right.
 \kern-\nulldelimiterspace} {\sqrt {{\epsilon _0}{\mu _0}} }}$. Without loss of generality the electric field is assumed polarized at the $y$ axis: ${\bf{E}} = {\left[ {0,E(t,x),0} \right]^T}$, while the magnetic induction field is parallel to the $z$ axis: ${\bf{B}} = {\left[ {0,0,B(t,x)} \right]^T}$. Hence, Maxwell's curl equations result (in the absence of charges and currents $\rho = J = 0$),
 
 \begin{eqnarray}
{\partial _x}E(t,x) + {\partial _t}B(t,x) = 0,
\label{eq:curl1}
 \\
{\partial _t}E(t,x) + {c^2}{\partial _x}B(t,x) = 0.
\label{eq:curl2}
\end{eqnarray}

It is additionally assumed that $r_{2}(t)>r_{1}(t)$ so that the mirrors never collide while obviously: $\left| {r'_{1}(t)} \right|<c$, $\left| {r'_{2}(t)} \right| < c$,  where prime indicates the time derivative. The boundary conditions for $x=r_{1}(t)$ and $x=r_{2}(t)$ are,

\begin{eqnarray}
 E(t,r_{1}(t)) - {r'_{1}(t)}B(t,r_{1}(t)) = 0,
\label{eq:bc1}
 \\
E(t,r_{2}(t)) - {r'_{2}(t)}B(t,r_{2}(t)) = 0,
\label{eq:bc2}
\end{eqnarray}

where the relativistic effects of the moving mirrors are taken into account \cite{costen}. It is worth noticing that the corrected relativistic boundary conditions arise from the fact that the integration over a small pill box at the moving boundary does not commute with the time derivative and an additional term related to the velocity of the boundary and the magnetic field has to be considered. Notice, that as expected if the mirrors are at rest the tangential electric field vanishes at the boundaries. The energy of the electromagnetic fields is,

\begin{equation}
e(t) = \frac{1}{2}\int_{r_{1}(t)}^{r_{2}(t)} {\left( {{\epsilon_0} {E(t,x)^2} + \frac{{{B(t,x)^2}}}{\mu_0}} \right)dx}. 
\label{eq:em_energy}
\end{equation}

For any time of observation, the electric and magnetic fields inside the mirrors have components that either propagate at the positive or the negative direction. We introduce the following wave functions in order to separate the positive and negative propagating transient signals,

\begin{eqnarray}
{\psi _1}(t,x) = E(t,x) + cB(t,x),
\label{eq:psi1}
 \\
{\psi _2}(t,x) = E(t,x) - cB(t,x).
\label{eq:psi2}
\end{eqnarray}

From (\ref{eq:curl1}) and (\ref{eq:curl2}), we find

\begin{eqnarray}
{\partial _t}{\psi _1}(t,x) + c{\partial _x}{\psi _1}(t,x) = 0,
\label{eq:psi1_w}
 \\
{\partial _t}{\psi _2}(t,x) - c{\partial _x}{\psi _2}(t,x) = 0.
\label{eq:psi2_w}
\end{eqnarray}

It is clear that (\ref{eq:psi1_w}) corresponds to the one-way wave equation with positive direction on the $x$ axis, while (\ref{eq:psi2_w}) corresponds to the one-way wave equation with negative direction at the $x$ axis. The boundary conditions for ${\psi _1}$ and ${\psi _2}$ at $x=r_{1}(t)$ and $x=r_{2}(t)$ are,

\begin{eqnarray}
{\psi _1}(t,r_{1}(t)) + \frac{{1 + {{r'_{1}(t)} \mathord{\left/
 {\vphantom {{r'_{1}(t)} c}} \right.
 \kern-\nulldelimiterspace} c}}}{{1 - {{r'_{1}(t)} \mathord{\left/
 {\vphantom {{r'_{1}(t)} c}} \right.
 \kern-\nulldelimiterspace} c}}}{\psi _2}(t,r_{1}(t)) = 0,
\label{eq:psi_bc1}
 \\
{\psi _2}(t,r_{2}(t)) + \frac{{1 - {{r'_{2}(t)} \mathord{\left/
 {\vphantom {{r'_{2}(t)} c}} \right.
 \kern-\nulldelimiterspace} c}}}{{1 + {{r'_{2}(t)} \mathord{\left/
 {\vphantom {{r'_{2}(t)} c}} \right.
 \kern-\nulldelimiterspace} c}}}{\psi _1}(t,r_{2}(t)) = 0.
\label{eq:psi_bc2}
\end{eqnarray}

Equivalently with (\ref{eq:em_energy}), the electric and magnetic energy in relation to ${\psi _1}$ and ${\psi _2}$ can be easily shown to be

\begin{equation}
e(t) = \frac{\epsilon_0}{4} \int_{r_{1}(t)}^{r_{2}(t)} {\left( {\psi _1(t,x)^2 + \psi _2(t,x)^2} \right)dx}.
\label{eq:em_energy_psi}
\end{equation}

The wave solutions (\ref{eq:psi1_w})-(\ref{eq:psi2_w}) indicate that the fields remain constant for the geometric lines of $t-x/c=$const and $t+x/c=$const at $r_{1}(t)<x<r_{2}(t)$. In fact these characteristics (rays) can be used in combination with the boundary conditions (\ref{eq:psi_bc1})-(\ref{eq:psi_bc2}) to find the solutions of the fields in the $(t,x)-$space (this method is known as the method of characteristics, see e.g., \cite{Garabedian,John}).

Suppose we are interested in the solution of the electric field at some time and space $(t,x)$, while the field has been created in the past (e.g., by an electric current that has been switched off). We are interested therefore in finding a solution of the D'Alembert type, where the field is known for a past time ${t_0} < t$. The electric field is

\begin{equation}
E(t,x) = \frac{1}{2}({\psi _1}(t,x) + {\psi _2}(t,x)).
\label{eq:E_psi}
\end{equation}

To find $E(t,x)$, one has to consider the two characteristics that pass through $(t,x)$ and propagate them backwards in time according to the rule that, upon reaching a mirror, they change direction of propagation, until they reach the line $t = {t_0}$, as seen in Fig. \ref{fig:fig_2}. Hence, $E(t,x)$ is given by

\begin{widetext}
\begin{equation}
E(t,x) = \frac{1}{2}\sum\limits_{\alpha  = 1}^2 {\left[ {{{( - 1)}^{{N_\alpha }}}\prod\limits_{n = 1}^{N_\alpha ^*} {\left( {\frac{{1 - {{{{r'_2}}(\theta _n^{(\alpha )})} \mathord{\left/
 {\vphantom {{{{r'_2}}(\theta _n^{(\alpha )})} c}} \right.
 \kern-\nulldelimiterspace} c}}}{{1 + {{{{r'_2}}(\theta _n^{(\alpha )})} \mathord{\left/
 {\vphantom {{{{r'_2}}(\theta _n^{(\alpha )})} c}} \right.
 \kern-\nulldelimiterspace} c}}}} \right)\prod\limits_{m = 1}^{{N_\alpha } - N_\alpha ^*} {\left( {\frac{{1 + {{{{r'_1}}(\tau _m^{(\alpha )})} \mathord{\left/
 {\vphantom {{{{r'_1}}(\tau _m^{(\alpha )})} c}} \right.
 \kern-\nulldelimiterspace} c}}}{{1 - {{{{r'_1}}(\tau _m^{(\alpha )})} \mathord{\left/
 {\vphantom {{{{r'_1}}(\tau _m^{(\alpha )})} c}} \right.
 \kern-\nulldelimiterspace} c}}}} \right){\psi _\alpha }({t_0},x_0^{(\alpha )})} } } \right]},
\label{eq:E_char}
\end{equation}
\end{widetext}

where $\alpha  = 1$ or $2$ and $x_0^{(1)}$, $x_0^{(2)}$ are the projection points of the characteristics at the $t={t_0}$ line (as seen in Fig. \ref{fig:fig_2}). $\theta_n^{(1)}$ and $\theta_n^{(2)}$ are the times the rays are reflected by the moving mirror at $x=r_{2}(t)$, $N_1^*$ and $N_2^*$ are the number of reflections from the moving mirror at $x=r_{2}(t)$ and $N_1$, $N_2$ are the total reflections. $\tau_m^{(1)}$ and $\tau_m^{(2)}$ are the times the rays are reflected by the moving mirror at $x=r_{1}(t)$, while $N_1-N_1^*$, $N_2-N_2^*$ are the number of reflections from the moving mirror at $x=r_{1}(t)$. If $N_{\alpha}^*=0$ or $N_{\alpha}-N_{\alpha}^*=0$ the associated multiplication is equal to $1$. Similar equation with (\ref{eq:E_char}) is found for the magnetic induction $B(t,x)$. From (\ref{eq:E_char}), we can define the Doppler factor:

\begin{widetext}
\begin{equation}
D({t_0},x_0^{(\alpha )};t) = \prod\limits_{n = 1}^{N_\alpha ^*} {\left( {\frac{{1 - {{{{r'_2}}(\theta _n^{(\alpha )})} \mathord{\left/
 {\vphantom {{{{r'_2}}(\theta _n^{(\alpha )})} c}} \right.
 \kern-\nulldelimiterspace} c}}}{{1 + {{{{r'_2}}(\theta _n^{(\alpha )})} \mathord{\left/
 {\vphantom {{{{r'_2}}(\theta _n^{(\alpha )})} c}} \right.
 \kern-\nulldelimiterspace} c}}}} \right)\prod\limits_{m = 1}^{{N_\alpha } - N_\alpha ^*} {\left( {\frac{{1 + {{{{r'_1}}(\tau _m^{(\alpha )})} \mathord{\left/
 {\vphantom {{{{r'_1}}(\tau _m^{(\alpha )})} c}} \right.
 \kern-\nulldelimiterspace} c}}}{{1 - {{{{r'_1}}(\tau _m^{(\alpha )})} \mathord{\left/
 {\vphantom {{{{r'_1}}(\tau _m^{(\alpha )})} c}} \right.
 \kern-\nulldelimiterspace} c}}}} \right)} } 
\label{eq:doppler}
\end{equation}
\end{widetext}

The Doppler factor (\ref{eq:doppler}) is clearly related to the multiplication of the ratios of the spatial distances between the characteristics before and after the multiple reflections from the moving mirrors, it is connected thus with the associated Jacobian: $D({t_0},x_0^{(\alpha)} ;t) = \left| {{{\partial x_0^{(\alpha)} } \mathord{\left/
 {\vphantom {{\partial x_0^{(\alpha)} } {\partial x}}} \right.
 \kern-\nulldelimiterspace} {\partial x}}} \right|$. The energy therefore can be calculated from the equation:
 
 \begin{figure}
\includegraphics{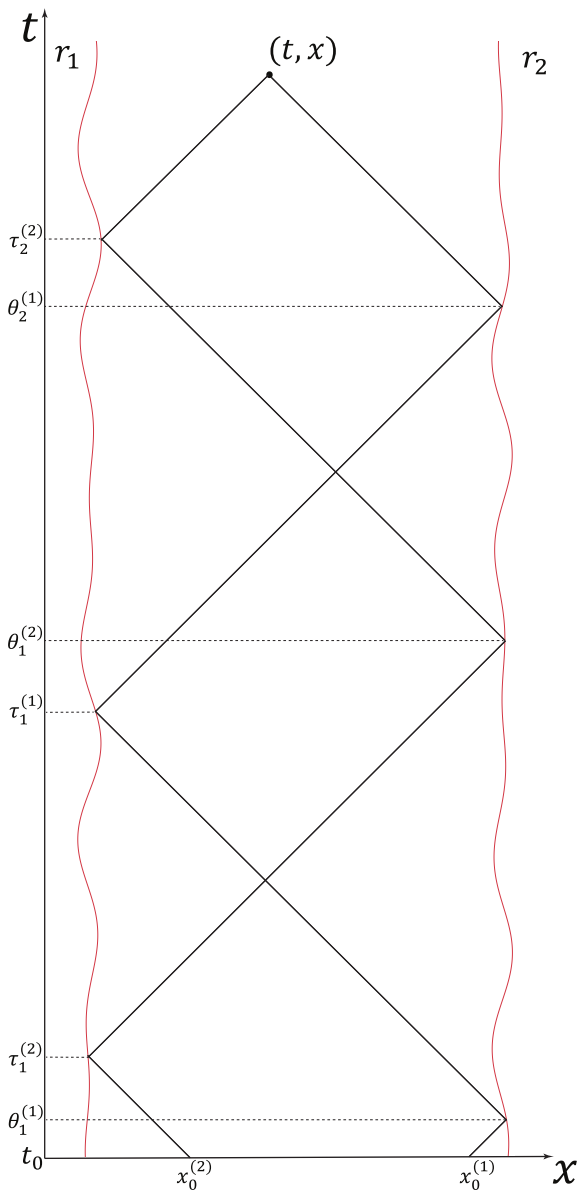}% Here is how to import EPS art
\caption{\label{fig:fig_2} Graphical illustration of the solution of the method of the characteristics on the $(t,x)$ space. Starting from the point of interest $(t,x)$, we follow backwards the lines of the characteristics, which reflect at the boundaries $r_1$ and $r_2$ until they intersect with $t=t_0$.}
\end{figure}
 
  \begin{equation}
e(t) = \frac{\epsilon_0}{4}\sum\limits_{\alpha  = 1}^2 {\left( {\int_{{r_1}({t_0})}^{{r_2}({t_0})} {D({t_0},x_0^{(\alpha )};t){\psi _\alpha }{{({t_0},x_0^{(\alpha )})}}^2dx_0^{(\alpha )}} } \right)}. 
\label{eq:energy_char}
\end{equation}

It is clear that the energy changes are dependent on the times of reflection $\theta_n^{(\alpha)}$ and $\tau_m^{(\alpha)}$ and the associated derivatives of $r_{2}(t)$ and $r_{1}(t)$, at these times respectively. Therefore, the displacements $r_{1}(t)$ and $r_{2}(t)$, may be engineered to either amplify or even attenuate the total energy.

In the special case where the mirrors travel in time with a constant velocity: $r'_{1}(t)=u_{1}$ and $r'_{2}(t)=u_{2}$, one may inspect the energy from (\ref{eq:energy_char}) for every two consecutive reflections. The energy is changing by a factor of $d={{(1 - {{{u_2}} \mathord{\left/
 {\vphantom {{{u_2}} c}} \right.
 \kern-\nulldelimiterspace} c})(1 + {{{u_1}} \mathord{\left/
 {\vphantom {{{u_1}} c}} \right.
 \kern-\nulldelimiterspace} c})} \mathord{\left/
 {\vphantom {{(1 - {{{u_2}} \mathord{\left/
 {\vphantom {{{u_2}} c}} \right.
 \kern-\nulldelimiterspace} c})(1 + {{{u_1}} \mathord{\left/
 {\vphantom {{{u_1}} c}} \right.
 \kern-\nulldelimiterspace} c})} {\left[ {(1 + {{{u_2}} \mathord{\left/
 {\vphantom {{{u_2}} c}} \right.
 \kern-\nulldelimiterspace} c})(1 - {{{u_1}} \mathord{\left/
 {\vphantom {{{u_1}} c}} \right.
 \kern-\nulldelimiterspace} c})} \right]}}} \right.
 \kern-\nulldelimiterspace} {\left[ {(1 + {{{u_2}} \mathord{\left/
 {\vphantom {{{u_2}} c}} \right.
 \kern-\nulldelimiterspace} c})(1 - {{{u_1}} \mathord{\left/
 {\vphantom {{{u_1}} c}} \right.
 \kern-\nulldelimiterspace} c})} \right]}}$. If ${u_\alpha } > 0$ then the corresponding mirror moves towards $\hat x$, while if ${u_\alpha } < 0$ it moves towards $-\hat x$. If the distance between the mirrors remains invariant then $d=1$ and the overall energy of the electromagnetic waves is constant despite the multiple Doppler reflections. If the mirrors approach (relativistically) each other then $d>1$, meaning that the energy is amplified. On the other hand, if the mirrors are moving away from each other, $d<1$ and the overall energy of the waves is decreased.
 
 Of course these observations are directly related to the known Doppler effects. Consider the change of the energy of a very narrow wave packet after two consecutive reflections. The temporal and the
spatial distances decrease by a factor of $1/d$, while the absolute values of the electric field and the magnetic induction will increase by $d$ times. Therefore, the integrand of the energy integral will increase
$d^2$ times, while the support of the integrand (i.e., the spatial width of the wave packet at $t$) will shrink by a factor of $1/d$. Hence, the energy of the wave packet after two reflections will be $d$ times greater than its energy before the reflections. 

These remarks are also valid for the temporally and spatially local effects of the characteristics when arbitrary functions of displacement $r_{1}(t)$ and $r_{2}(t)$ are considered. If a characteristic follows a path that reflects on the boundaries at times $\theta_n$ and $\tau_m$ such that the characteristic is squeezed inside the cavity, its local energy density will increase, while if the reflections dilute the characteristics then the local energy density of the wave decreases. This in fact leads to positions of the cavity that can maximize and amplify the electric field distribution and to positions that minimize and vanish the field.

\begin{figure}
\includegraphics{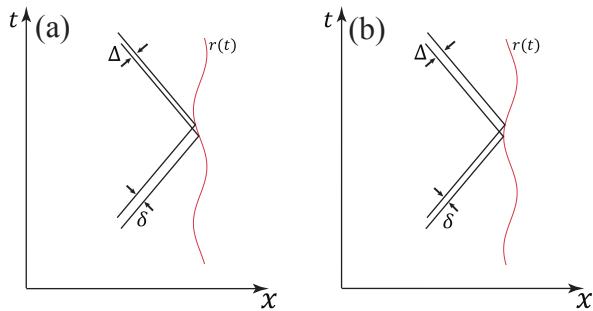}% Here is how to import EPS art
\caption{\label{fig:fig_3}Diagram of neighbor characteristics at the $(t,x)-$space. $\delta$ is the distance between the characteristics before reflection and $\Delta$ the distance after. (a) The movement of the boundary squeezes the characteristics ($\Delta < \delta$). (b) The movement of the boundary loosens the characteristics ($\Delta > \delta$).}
\end{figure}

These claims are evident from a geometrical point of view. Assume two neighbor characteristics before and after the reflection of a boundary to the right of the characteristics in the $(t,x)$ space, as seen in Fig. \ref{fig:fig_3}. Before the reflection, their distance is noted as $\delta$ and after the reflection $\Delta$. It is easily shown that $\mathop {\lim }\limits_{(\delta ,\Delta)  \to 0} \left( \tfrac{\delta}{\Delta}  \right) = 
\tfrac{1 - r'(t_{ref})}{1 + r'(t_{ref})}$ \footnote{The equation has opposite signs for the velocity, if the characteristics were reflected from the left instead of the right as seen in Fig. \ref{fig:fig_3}}. Therefore, depending on the value of $r'(t_{ref})$ at the time of reflection $t_{ref}$, the characteristics can be squeezed meaning that the distances of neighbor characteristics decrease leading to the concentration of fields, or the characteristics sparse, meaning that the distances between two neighbor characteristics increase leading to the dispersion of the rays and therefore the decrease of the field local energy density.

In relation to (\ref{eq:em_energy_psi}), it is clear that the energy increases whenever the transients are in contact with the moving boundaries and the plates move toward the incident wave. The frequencies of the incident wave are augmented according to the Doppler shift. This is why these phenomena are both temporally and spatially local and may lead to the frequency filtering and growth of the electromagnetic transients. The mechanical energy of the moving plates is converted into the energy of the wave for $r'_{1}(t)>0$ and $r'_{2}(t)<0$, while the reverse occurs if the plates move away from each other.

This amplification mechanism is somewhat analogue with the parametric amplification phenomena in time-periodic optical media \cite{Cassedy1,Cassedy2,koutserimpas1} and the compression of lines of forces as described by Pendry \emph{et al.} \cite{pendry1}. Contradictory to these studies, the amplification mechanism analyzed here which is based on the local Doppler effects of the characteristics, doesn't require the modulation of a bulk medium, neither very fast modulations and therefore could be easier implemented.

\section{\label{sec:periodic_boundaries}Periodically moving boundaries}

\subsection{\label{subsec:periodic_boundaries_gen} Periodic characteristics: Amplification and attenuation mechanisms}

The analysis of the previous Section is general and for arbitrary smooth time-profiles of $r_{1}(t)$ and $r_{2}(t)$. The times of reflection $\tau$ and $\theta$ can therefore take any values that correspond to the associate Doppler factor resulting in the variation of the energy of the electromagnetic transients, depending on the time-derivatives of the boundary functions at the associated reflection times. In this Section, the boundary profiles are considered periodic: ${r_\alpha }(t) = {f_\alpha }(t)$, with ${f_\alpha }(t)={f_\alpha }(t+T)$ and $\tfrac{1}{T}\int_0^T {({f_2 }(t)-{f_1 }(t))}  = L$, where $T = \tfrac{2\pi}{\Omega}$ is the period and $L$ is the mean distance between the plates. In such systems the position of the characteristics for every $t=t_{0}+nT_{char}$, $n \in \mathbb{Z}^\ge$ is of interest. Notice that the period of the characteristics $T_{char}$ may not be equal with the period of modulation $T$ (often it is $T_{char}= 2L/c$ and an integer multiple of $T$). 

If a discrete set of points exists $\chi _{1}<\cdots<\chi _{i}< \cdots <\chi _{M}$ where the characteristics maintain the same positions for every snapshot at $t=t_{0}+nT_{char}$, then due to the periodicity of the boundaries we have $D({t_0},{\chi _i};{t_0} + nT_{char}) = D{({t_0},{\chi _i};{t_0} + T_{char})^n}$. If the Doppler factor is chosen such that results in $D{({t_0},{\chi _a};{t_0} + T_{char})}>1$, the overall Doppler effect results in condensating the neighbor characteristics, asymptotically leading to $\mathop {\lim }\limits_{n  \to \infty} D{({t_0},{\chi _a};{t_0} + T_{char})^n}=\infty$. The characteristics converge at the periodic attractive characteristic which passes through $(t_{0},\chi_{a})$. Analogously, if $D{({t_0},{\chi _r};{t_0} + T_{char})}<1$, the overall Doppler effect over the $T_{char}$ period results in the repulsion of the characteristics from the associated point $\chi_{r}$ (the characteristics on the left side of $\chi_{r}$ are attracted by the attractive point $\chi_{r-1}$, while those on the right side are attracted by $\chi_{r+1}$) and asymptotically $\mathop {\lim }\limits_{n  \to \infty} D{({t_0},{\chi _r};{t_0} + T_{char})^n}=0$. The characteristic therefore diverges from the associated repulsive periodic characteristic which passes through $(t_{0},\chi_{r})$ meaning that the field is reshaped and asymptotically reaches zero at the repelling point after the course of many periods. For any square integrable $\psi_{1}(t_{0},x)$ and $\psi_{2}(t_{0},x)$, the solutions converge to generalized functions that are concentrated on the periodic attractive characteristics and $e(nT_{char}) \sim \exp (\kappa T_{char})$ with $\kappa  = \tfrac{n}{T_{char}} \ln \left[ {D({t_0},{\chi _a};{t_0} + T_{char})} \right]$. This physically means that the wave solutions get asymptotically synchronized so that there is a net increase in energy through each period.

In the case of $D{({t_0},{\chi _n};{t_0} + T_{char})}=1$ one has to identify the conditions for the neighbor points $\chi _{n-1}$ and $\chi _{n+1}$ to get a better picture of the movement of the characteristics (it has been proven in \cite{Gonzalez} that if certain conditions are met polynomial field growth may occur). 

These phenomena contain similarities with the kinematic mode locking laser configurations, where a laser cavity with a moving mirror can induce mode locking due to the interplay between Doppler frequency shift by the moving mirror, frequency filtering by the gain medium and spectral broadening by Kerr nonlinearities \cite{smith1,sterke1}.

\subsection{\label{subsec:periodic_boundaries_cs1}The case of ${r_1}(t) =  - \Delta L\sin (\Omega t)$ and $r_{2}(t)= L + \Delta L\sin (\Omega t)$}

In the special case where: ${r_1}(t) =  - \Delta L\sin (\Omega t)$ and $r_{2}(t)= L + \Delta L\sin (\Omega t)$ while $t_{0}=0$, the characteristics can result in discrete points $\chi_{i}$ under specific conditions. Namely, if the period of the vibrations of the plates satisfies
 
  \begin{equation}
T=\frac{L}{Nc},
\label{eq:period_condition}
\end{equation}
 
 where $N \in \mathbb{Z}^>$. A total of $M=4N+1$ discrete isodistanced $\chi_{i}$ points are found, where
 
   \begin{equation}
\chi_{i}=\frac{(i-1)L}{4N}, \quad  i=1,\cdots,4N+1.
\label{eq:period_points}
\end{equation}

For $\Delta L>0$ (or $\Delta L <0$) the corresponding $\chi_{i}$ where $i$ is even (or odd) is a point in which attractive periodic characteristics pass through at every time slot $nT_{char}$, while for an odd (or even) $i$ the characteristics that pass through are repulsive. The value of $M$ which is set by the vibration movement of the plates has an upper limit

  \begin{equation}
M < \frac{2L}{\pi \Delta L}+1.
\label{eq:period_limit}
\end{equation}

The number of periodic characteristics and therefore peaks (or short pulses) is limited by the fact that the velocity of the mechanical vibrations can not exceed that of light. The period of the periodic characteristics is equal to $T_{char}= 2L/c$ (or equivalently $T_{char}=\tfrac{M-1}{2}T$).

\section{\label{sec:fdtd}FDTD simulations}

\begin{figure*}
\includegraphics{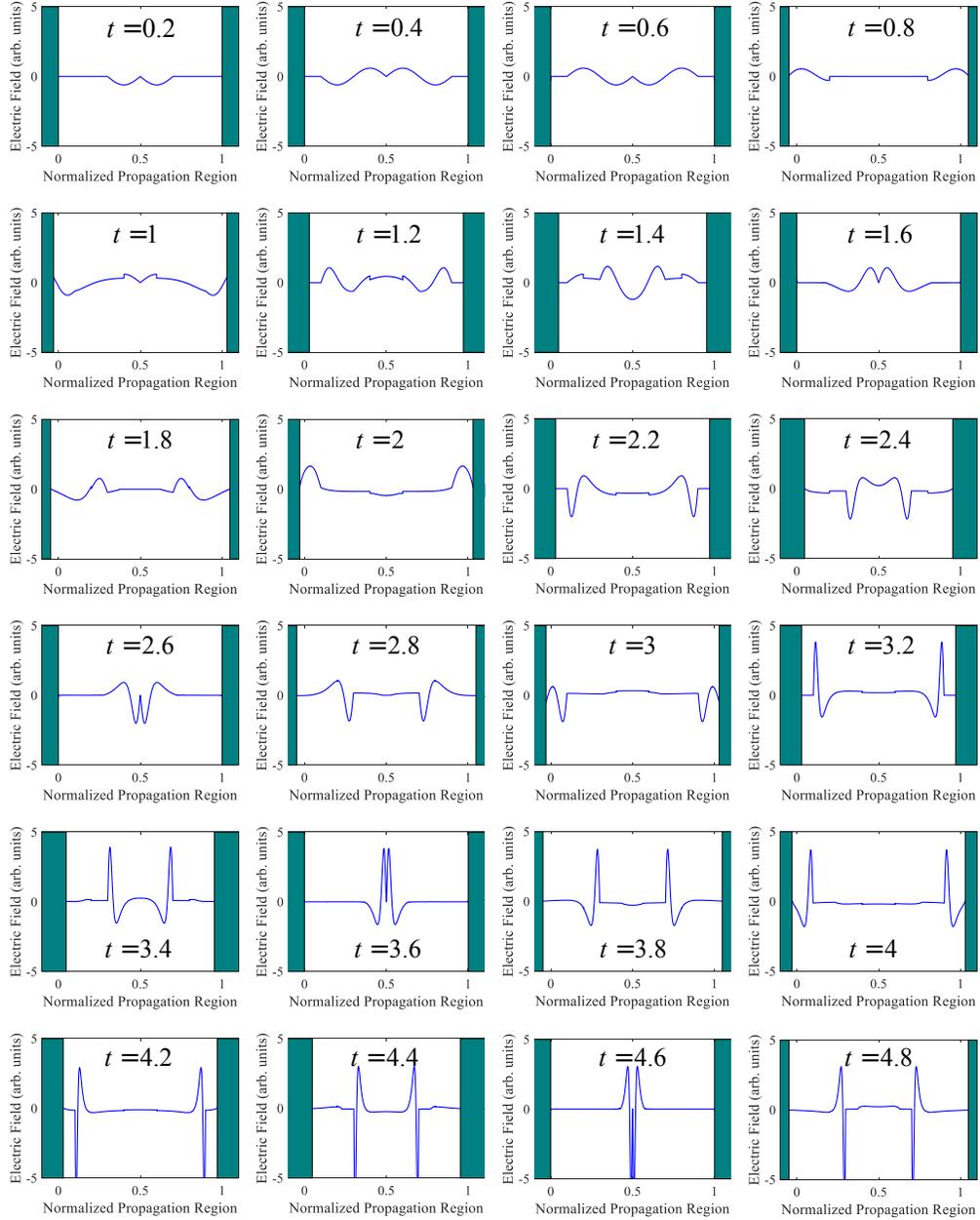}% Here is how to import EPS art
\caption{\label{fig:fig_4} FDTD chronophotographs of the electric field transients with periodically moving boundaries for $c=1$, $L=1$, $\Delta L=0.05$ and $T = 1$. The snapshots of the FDTD simulation are taken for every $\Delta t =0.2$. The associated movie of the electric field oscillations of this case depicted is the first movie in the Supplementary Material.}
\end{figure*}
 
 \begin{figure*}
\includegraphics{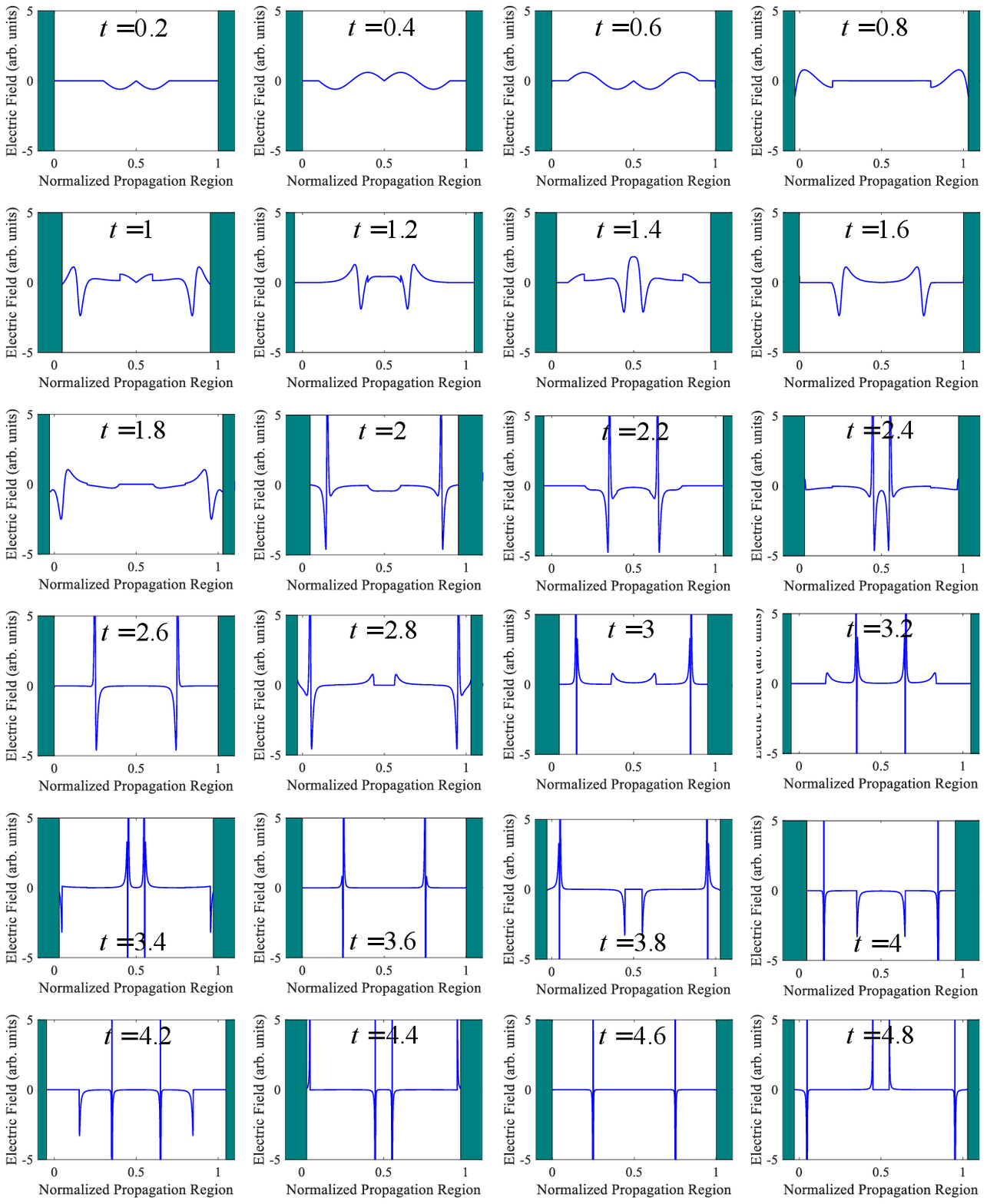}% Here is how to import EPS art
\caption{\label{fig:fig_5} FDTD chronophotographs of the electric field transients with periodically moving boundaries for $c=1$, $L=1$, $\Delta L=0.05$ and $T = 0.5$. The snapshots of the FDTD simulation are taken for every $\Delta t =0.2$. The associated movie of the electric field oscillations of this case depicted is the second movie in the Supplementary Material.}
\end{figure*}
 
 \begin{figure}
\includegraphics{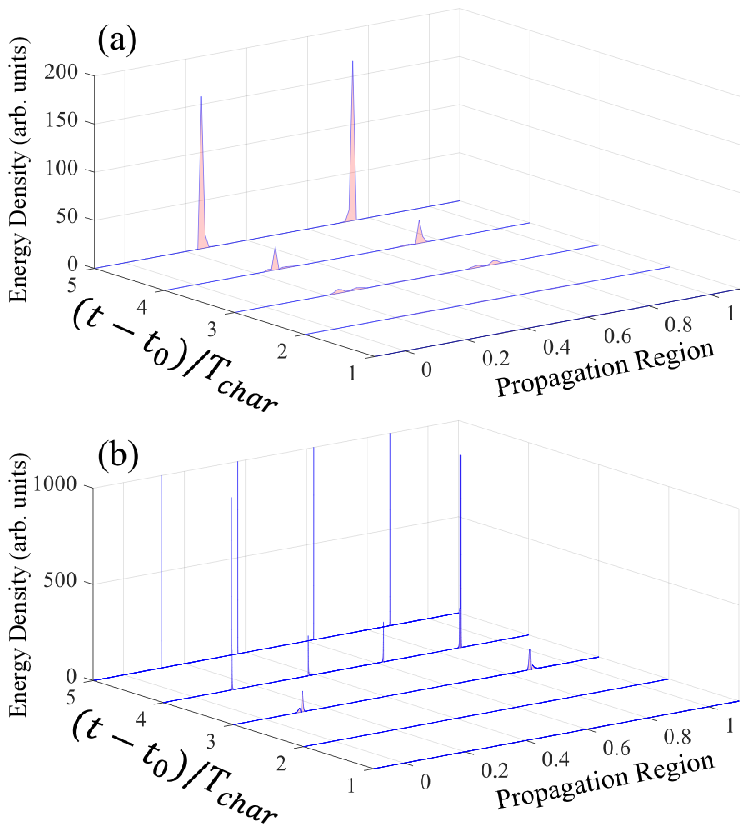}% Here is how to import EPS art
\caption{\label{fig:fig_6} FDTD chronophotographs of the energy density with periodically moving boundaries. The model assumes $c=1$, $L=1$ and $\Delta L=0.05$. The energy density distribution is exhibited over the course of 5 periods $T_{char}=2$ for (a) $T = 1$ and (b) $T = 0.5$. The energy density distributions form delta-like distributions at the attractive periodic points as theoretically predicted in Section (\ref{subsec:periodic_boundaries_cs1}). The associated movies of the electric field oscillations of the two cases depicted are found in the Supplementary Material.}
\end{figure}

For the numerical verification of the presented analysis a homemade 1D FDTD full-wave model was implemented, (see e.g., \cite{Taflove,harfoush}). The FDTD method relies on the discretization of time and space and therefore the only approximations are in the associated derivatives of space and time. 

For the purposes of illustration of the phenomena we normalize $c=1$ and $L=1$, while we fix $\Delta L =0.05$. The region is initially excited by a point current source at the middle of the computation region from $t=0$ until $t=0.6$. The current creates transient waves that propagate forward and backwards that reflect at the boundaries. At exactly $t=0.6$, the current is switched off and the boundaries start to move leading to a spatial filtering and amplification of the transients related to the Doppler factor as described in the previous sections. 

In Fig. \ref{fig:fig_4}, snapshots of the electric field oscillations, for every $\Delta t =0.2$ are exhibited. The period of the moving boundaries is set at $T=1$, which corresponds to $M=5$. As we see in the related FDTD chonophotographs of the FDTD calculations, the electric field which is initially created by an oscillating current, is reflected from the left and the right, while the boundaries are in movement. Over the course of the simulation, the transient signal is reshaped into two sharp peaks (as predicted by the characteristics analysis) which are synchronized and are reflected at every time slot that the boundaries develop maximum velocity towards each other. It is clear that the modulation frequency is matched with the cavity's length. This is in fact in agreement with the FM mode locking theory (see Chapter 27 in \cite{Siegman}). In contrast to the FM mode locking theory, the pulses of the initial excitation of the current at the middle of the mirrors form antisymmetric field shapes. The FM mode locking theory predicts that the frequency chirp of the two pulses should be of opposite signs, but this is not the case in our analysis, where the two pulses have in fact the same frequency chirp.

Similarly, in Fig. \ref{fig:fig_5} again the snapshots of the electric field are taken for every $\Delta t =0.2$, while the period of the moving boundaries is set at $T=1/2$, which corresponds to $M=9$. The FDTD chronophotographs show again the generated transient signal by the current while it is reflected by the moving boundaries. The condition of operation satisfy this time four peaks. Hence we see, as time passes that the transient signal is reshaped into four sharp peaks. Again, the boundaries are synchronized and develop their maximum velocity towards each other at the specific time slots when these wave packets reflect at the boundaries. The initial field discontinuity as found e.g., in Figs. \ref{fig:fig_4}-\ref{fig:fig_5} for $t=0.8$, is carried from the discontinuity of the velocities of the boundaries at $t=t_{0}$. In fact, under the configuration described in Fig. \ref{fig:fig_5} two out of four attractive characteristics amplify and concentrate the transients created by the discontinuity of the velocity of the mirrors, creating a pair of two positive half-cycle pulses that propagate in the cavity. The abrupt change of the velocity which is introduced by the mirrors is translated to a sudden change of the boundary conditions which actively reshape the signal. Such abrupt change of velocity is acceptable since a sudden change of the applied mechanical forces is allowed (it is of course, the displacement that has to be continuous).

In Fig. \ref{fig:fig_6}, the energy density distributions as computed by the FDTD model are shown over the course of 5 periods $T_{char}=2$, it is clear that since the computations are for more time than in Figs. \ref{fig:fig_4}-\ref{fig:fig_5}, the delta-like peaks of the density energy are formulated. At Fig. \ref{fig:fig_6}a, the period is set $T=1$, which means that over the course of multiple $T_{char}$ periods the characteristics concentrate at the attractive periodic points which are (in full agreement with the theoretical analysis (\ref{eq:period_points})) at $\chi_{2}=0.25$ and $\chi_{4}=0.75$. Fig \ref{fig:fig_6}a shows clearly that the energy density is amplified and concentrated at these points as we observe the energy density for each period $T_{char}$, leading to delta-like distributions of energy. Analogously, at Fig \ref{fig:fig_6}b, the energy density distributions for $T=0.5$ are exhibited using the FDTD model. In this case, (again in full agreement with the theoretical analysis (\ref{eq:period_points})) the energy concentrates at the attractive periodic points: $\chi_{2}=0.125$, $\chi_{4}=0.375$, $\chi_{6}=0.625$ and $\chi_{8}=0.875$. 

The FDTD results verify that the characteristics indeed concentrate at the periodic attractive points, leading to the creation of ultrashort pulses that are exponentially amplifying due to their synchronization with the moving boundaries. The associated videos of the FDTD simulations of Figs \ref{fig:fig_4}-\ref{fig:fig_5} can be found in the Supplementary Material \footnote{Supplementary material with two videos of the electric field oscillations for: (a) $T=1$ and (b) $T=0.5$.}.

\section{\label{sec:discussion}Discussion}

The interactions of electromagnetic transients trapped inside two moving perfect mirrors were studied using the method of characteristics and FDTD simulations. The wave propagation was found to be dominantly affected by the temporally and spatially local Doppler phenomena. The Doppler effect can filter and amplify or attenuate the electromagnetic signal, depending on the movement of the mirrors at the point $(t,x)$ of reflection in respect to the direction of the incident signal. To understand further the active mechanism, we considered the special case of the periodically vibrating mirrors that under the condition of having a cavity period ($L/c$) equal to integer multiples of the period of the mechanical motion, periodic characteristics can exist. Such periodic characteristics can be either attractive or repulsive, squeezing the transients or diluting them in the corresponding periodic characteristic. Therefore over the course of time the electric field forms delta-like field distributions at the points of the attractive periodic characteristics. For illustration purposes, we performed FDTD simulations, which verified our theoretical findings. 

The aforementioned gain mechanism resembles the so called kinematic mode locking first described by Smith in \cite{smith1}, where the Doppler frequency shift by the moving mirror in combination with the filtering by the gain medium and the spectral broadening by the Kerr effect can establish a phase-locked laser \cite{sterke1}. More recent laser configurations, substitute the Doppler frequency shift with acousto-optic effects \cite{Noronen1,Woodward1,Majewski1} and electro-optic effects \cite{Kuizenga1,Kuizenga2,Siegman,kovanis} and attain mode locking conditions in relation to the fundamental frequency and the modulation frequencies that are similar to the one of our analysis (\ref{eq:period_condition}). In contrast to the laser configurations, our study is solely based on the Doppler and relativistic effects and the intriguing dynamics they induce in the trapped electromagnetic field.

Additionally, the proposed gain mechanism is different than the one that has been recently studied in the context of time-periodic optical media (see e.g. \cite{pendry1,koutserimpas1}). That is in the sense that the gain mechanism described in the present article doesn't come from a complex eigenvalue related to the eigensolutions of the Floquet theory of the wave operator in periodic media; it is generated by the local Doppler effects which reshape the electromagnetic signal. The energy growth results from the compression of the pulse and its magnification in an arbitrarily small region near the attractive periodic characteristic. This means that amplification by temporal modulation of a homogeneous medium will affect the whole  field simultaneously in the same way (if one point of the field amplifies then every point of the field will amplify), while in our proposition the active filtering by the moving boundaries will only amplify the points at the attractive periodic characteristic (resulting in pulses).

Although the object of this study was not a description of suitable agencies for realizing the assumed mechanical configuration, some remarks as to how such structure can be engineered seem to be in order. Possible implementations may incorporate technologies involving actuators \cite{grosso1}. For instance, it has been shown that the electrothermal transduction of graphene-polymer resonators may result in resonant frequencies within the audio range \cite{Al1}. Of course the velocities generated may be low and multiple reflections should pass in order to start exhibiting the peaks. Other possible implementations in the nanoscale may include micromechanical resonators or microscale acoustofluidics that can reach to higher mechanical frequency modulation in the magnitude of MHz and GHz and decrease the required $L$ for the periodic attractive characteristics condition to the magnitude of a few meters, as well as the required time spend for the same absolute value amplification \cite{Groeblacher,Anetsberger,Friend}. For example, using the laser configurations simulated in \cite{kovanis}, a $L=1.25$ mm cavity with a a mechanical motion $\Omega = 151 \cdot {10^{10}}$ rad/sec and $\Delta L = 15$ nm will amplify three times its initial maximum amplitude in $5.375 \cdot {10^{-8}}$ sec. 

Additionally, notice that in our analysis we have assumed perfect mirrors of zero losses. This is applicable at the microwave range. At the optical range losses can no longer be ignored. Therefore, a trade off exists; on one hand at low electromagnetic frequencies the model should match the experiments but should require multiple reflections to exhibit the asymptotic singularities, while on the other hand at higher electromagnetic frequencies the velocities can be substantial enough to quicker produce the predicted peaks but losses can no longer be neglected which should result in an additional filtering and attenuation of the wave signal.

Nevertheless, the presented analysis enriches the established literature in electromagnetic wave propagation in dynamically changing structures and systems as well as investigates the Doppler effect as an interesting gain mechanism which may be easier to implement, in comparison with the time-modulation of a bulk medium. We further establish an alternative method for the creation of ultrashort pulses which can have applications in the microwave and optical regimes and shares conceptual similarities with the mode locking theory.

\begin{acknowledgments}
TTK is supported by a Post.Doc Mobility Fellowship from the Swiss National Science Foundation (SNSF), Grant No. 203176. CV is partially supported by Nazarbayev University Faculty Development Competitive Research Grant No. 021220FD4051 (``Optimal design of photonic and quantum metamaterials'').
\end{acknowledgments}

%\appendix

%\section{Appendixes}

\bibliography{bibliography1}% Produces the bibliography via BibTeX.

\providecommand{\noopsort}[1]{}\providecommand{\singleletter}[1]{#1}%
\begin{thebibliography}{44}%
\makeatletter
\providecommand \@ifxundefined [1]{%
 \@ifx{#1\undefined}
}%
\providecommand \@ifnum [1]{%
 \ifnum #1\expandafter \@firstoftwo
 \else \expandafter \@secondoftwo
 \fi
}%
\providecommand \@ifx [1]{%
 \ifx #1\expandafter \@firstoftwo
 \else \expandafter \@secondoftwo
 \fi
}%
\providecommand \natexlab [1]{#1}%
\providecommand \enquote  [1]{``#1''}%
\providecommand \bibnamefont  [1]{#1}%
\providecommand \bibfnamefont [1]{#1}%
\providecommand \citenamefont [1]{#1}%
\providecommand \href@noop [0]{\@secondoftwo}%
\providecommand \href [0]{\begingroup \@sanitize@url \@href}%
\providecommand \@href[1]{\@@startlink{#1}\@@href}%
\providecommand \@@href[1]{\endgroup#1\@@endlink}%
\providecommand \@sanitize@url [0]{\catcode `\\12\catcode `\$12\catcode
  `\&12\catcode `\#12\catcode `\^12\catcode `\_12\catcode `\%12\relax}%
\providecommand \@@startlink[1]{}%
\providecommand \@@endlink[0]{}%
\providecommand \url  [0]{\begingroup\@sanitize@url \@url }%
\providecommand \@url [1]{\endgroup\@href {#1}{\urlprefix }}%
\providecommand \urlprefix  [0]{URL }%
\providecommand \Eprint [0]{\href }%
\providecommand \doibase [0]{https://doi.org/}%
\providecommand \selectlanguage [0]{\@gobble}%
\providecommand \bibinfo  [0]{\@secondoftwo}%
\providecommand \bibfield  [0]{\@secondoftwo}%
\providecommand \translation [1]{[#1]}%
\providecommand \BibitemOpen [0]{}%
\providecommand \bibitemStop [0]{}%
\providecommand \bibitemNoStop [0]{.\EOS\space}%
\providecommand \EOS [0]{\spacefactor3000\relax}%
\providecommand \BibitemShut  [1]{\csname bibitem#1\endcsname}%
\let\auto@bib@innerbib\@empty
%</preamble>
\bibitem [{\citenamefont {Rayleigh}(1902)}]{rayleigh}%
  \BibitemOpen
  \bibfield  {author} {\bibinfo {author} {\bibfnamefont {L.}~\bibnamefont
  {Rayleigh}},\ }\bibfield  {title} {\bibinfo {title} {On the pressure of
  vibrations},\ }\href@noop {} {\bibfield  {journal} {\bibinfo  {journal}
  {Phil.\ Mag.}\ }\textbf {\bibinfo {volume} {3}},\ \bibinfo {pages} {338}
  (\bibinfo {year} {1902})}\BibitemShut {NoStop}%
\bibitem [{\citenamefont {Larmor}(1912)}]{larmor}%
  \BibitemOpen
  \bibfield  {author} {\bibinfo {author} {\bibfnamefont {J.}~\bibnamefont
  {Larmor}},\ }\bibfield  {title} {\bibinfo {title} {On the dynamics of
  radiation},\ }\href@noop {} {\bibfield  {journal} {\bibinfo  {journal}
  {Proceedings of the Fifth International Congress of Mathematicians,
  Cambridge}\ }\textbf {\bibinfo {volume} {1}},\ \bibinfo {pages} {197}
  (\bibinfo {year} {1912})}\BibitemShut {NoStop}%
\bibitem [{\citenamefont {Havelock}(1924)}]{havelock}%
  \BibitemOpen
  \bibfield  {author} {\bibinfo {author} {\bibfnamefont {T.~H.}\ \bibnamefont
  {Havelock}},\ }\bibfield  {title} {\bibinfo {title} {Some dynamical
  illustrations of the pressure of radiation and of adiabatic invariance},\
  }\href@noop {} {\bibfield  {journal} {\bibinfo  {journal} {Phil.\ Mag.}\
  }\textbf {\bibinfo {volume} {47}},\ \bibinfo {pages} {754} (\bibinfo {year}
  {1924})}\BibitemShut {NoStop}%
\bibitem [{\citenamefont {Nicolai}(1925)}]{nicolai}%
  \BibitemOpen
  \bibfield  {author} {\bibinfo {author} {\bibfnamefont {E.~L.}\ \bibnamefont
  {Nicolai}},\ }\bibfield  {title} {\bibinfo {title} {On a dynamical
  illustration of the pressure of radiation},\ }\href@noop {} {\bibfield
  {journal} {\bibinfo  {journal} {Phil.\ Mag.}\ }\textbf {\bibinfo {volume}
  {49}},\ \bibinfo {pages} {171} (\bibinfo {year} {1925})}\BibitemShut
  {NoStop}%
\bibitem [{\citenamefont {Minkowski}(1909)}]{minkowski}%
  \BibitemOpen
  \bibfield  {author} {\bibinfo {author} {\bibfnamefont {H.}~\bibnamefont
  {Minkowski}},\ }\bibfield  {title} {\bibinfo {title} {Raum und zeit},\
  }\href@noop {} {\bibfield  {journal} {\bibinfo  {journal} {Physikalische
  Zeitschrift}\ }\textbf {\bibinfo {volume} {10}},\ \bibinfo {pages} {104}
  (\bibinfo {year} {1909})}\BibitemShut {NoStop}%
\bibitem [{\citenamefont {Balazs}(1961)}]{balazs}%
  \BibitemOpen
  \bibfield  {author} {\bibinfo {author} {\bibfnamefont {N.}~\bibnamefont
  {Balazs}},\ }\bibfield  {title} {\bibinfo {title} {On the solution of the
  wave equation with moving boundaries},\ }\href@noop {} {\bibfield  {journal}
  {\bibinfo  {journal} {J. Math. Anal. Appl.}\ }\textbf {\bibinfo {volume}
  {3}},\ \bibinfo {pages} {472} (\bibinfo {year} {1961})}\BibitemShut {NoStop}%
\bibitem [{\citenamefont {Cooper}(1980)}]{cooper1}%
  \BibitemOpen
  \bibfield  {author} {\bibinfo {author} {\bibfnamefont {J.}~\bibnamefont
  {Cooper}},\ }\bibfield  {title} {\bibinfo {title} {Scattering of
  electromagnetic fields by a moving boundary: The one-dimensional case},\
  }\href@noop {} {\bibfield  {journal} {\bibinfo  {journal} {IEEE Trans.
  Antennas Propag.}\ }\textbf {\bibinfo {volume} {6}},\ \bibinfo {pages} {791}
  (\bibinfo {year} {1980})}\BibitemShut {NoStop}%
\bibitem [{\citenamefont {Bladel}\ and\ \citenamefont {Zutter}(1981)}]{bladel}%
  \BibitemOpen
  \bibfield  {author} {\bibinfo {author} {\bibfnamefont {J.~V.}\ \bibnamefont
  {Bladel}}\ and\ \bibinfo {author} {\bibfnamefont {D.~D.}\ \bibnamefont
  {Zutter}},\ }\bibfield  {title} {\bibinfo {title} {Reflections from linearly
  vibrating objects: Plane mirror at normal incidence},\ }\href@noop {}
  {\bibfield  {journal} {\bibinfo  {journal} {IEEE Trans. Antennas Propag.}\
  }\textbf {\bibinfo {volume} {29}},\ \bibinfo {pages} {629} (\bibinfo {year}
  {1981})}\BibitemShut {NoStop}%
\bibitem [{\citenamefont {Cooper}(1993)}]{cooper2}%
  \BibitemOpen
  \bibfield  {author} {\bibinfo {author} {\bibfnamefont {J.}~\bibnamefont
  {Cooper}},\ }\bibfield  {title} {\bibinfo {title} {Long-time behavior and
  energy growth for electromagnetic waves reflected by a moving boundary},\
  }\href@noop {} {\bibfield  {journal} {\bibinfo  {journal} {IEEE Trans.
  Antennas Propag.}\ }\textbf {\bibinfo {volume} {41}},\ \bibinfo {pages}
  {1365} (\bibinfo {year} {1993})}\BibitemShut {NoStop}%
\bibitem [{\citenamefont {de~la Llave}\ and\ \citenamefont
  {Petrov}(1999)}]{rafael1}%
  \BibitemOpen
  \bibfield  {author} {\bibinfo {author} {\bibfnamefont {R.}~\bibnamefont
  {de~la Llave}}\ and\ \bibinfo {author} {\bibfnamefont {N.~P.}\ \bibnamefont
  {Petrov}},\ }\bibfield  {title} {\bibinfo {title} {Theory of circle maps and
  the problem of one-dimensional optical resonator with a periodically moving
  wall},\ }\href@noop {} {\bibfield  {journal} {\bibinfo  {journal} {Phys. Rev.
  E}\ }\textbf {\bibinfo {volume} {59}},\ \bibinfo {pages} {6637} (\bibinfo
  {year} {1999})}\BibitemShut {NoStop}%
\bibitem [{\citenamefont {Petrov}\ \emph {et~al.}(2003)\citenamefont {Petrov},
  \citenamefont {de~la Llave},\ and\ \citenamefont {Vano}}]{rafael2}%
  \BibitemOpen
  \bibfield  {author} {\bibinfo {author} {\bibfnamefont {N.~P.}\ \bibnamefont
  {Petrov}}, \bibinfo {author} {\bibfnamefont {R.}~\bibnamefont {de~la
  Llave}},\ and\ \bibinfo {author} {\bibfnamefont {J.~A.}\ \bibnamefont
  {Vano}},\ }\bibfield  {title} {\bibinfo {title} {Torus maps and the problem
  of a one-dimensional optical resonator with a quasiperiodically moving
  wall},\ }\href@noop {} {\bibfield  {journal} {\bibinfo  {journal} {Physica
  D}\ }\textbf {\bibinfo {volume} {180}},\ \bibinfo {pages} {140} (\bibinfo
  {year} {2003})}\BibitemShut {NoStop}%
\bibitem [{\citenamefont {Dalvit}\ \emph {et~al.}(2011)\citenamefont {Dalvit},
  \citenamefont {Millonni}, \citenamefont {Roberts},\ and\ \citenamefont
  {Rosa}}]{Dalvit1}%
  \BibitemOpen
  \bibfield  {author} {\bibinfo {author} {\bibfnamefont {D.}~\bibnamefont
  {Dalvit}}, \bibinfo {author} {\bibfnamefont {P.}~\bibnamefont {Millonni}},
  \bibinfo {author} {\bibfnamefont {D.}~\bibnamefont {Roberts}},\ and\ \bibinfo
  {author} {\bibfnamefont {F.}~\bibnamefont {Rosa}},\ }\href@noop {} {\emph
  {\bibinfo {title} {Casimir Physics}}}\ (\bibinfo  {publisher} {Springer},\
  \bibinfo {year} {2011})\BibitemShut {NoStop}%
\bibitem [{\citenamefont {Moore}(1970)}]{Moore}%
  \BibitemOpen
  \bibfield  {author} {\bibinfo {author} {\bibfnamefont {G.~T.}\ \bibnamefont
  {Moore}},\ }\bibfield  {title} {\bibinfo {title} {Quantum theory of the
  electromagnetic field in a variable‐length one‐dimensional cavity},\
  }\href@noop {} {\bibfield  {journal} {\bibinfo  {journal} {J. Math. Phys.}\
  }\textbf {\bibinfo {volume} {11}},\ \bibinfo {pages} {2679} (\bibinfo {year}
  {1970})}\BibitemShut {NoStop}%
\bibitem [{\citenamefont {Ji}\ \emph {et~al.}(1998)\citenamefont {Ji},
  \citenamefont {Jung},\ and\ \citenamefont {Soh}}]{Ji1}%
  \BibitemOpen
  \bibfield  {author} {\bibinfo {author} {\bibfnamefont {J.-Y.}\ \bibnamefont
  {Ji}}, \bibinfo {author} {\bibfnamefont {H.-H.}\ \bibnamefont {Jung}},\ and\
  \bibinfo {author} {\bibfnamefont {K.-S.}\ \bibnamefont {Soh}},\ }\bibfield
  {title} {\bibinfo {title} {Interference phenomena in the photon production
  between two oscillating walls},\ }\href@noop {} {\bibfield  {journal}
  {\bibinfo  {journal} {Phys. Rev. A}\ }\textbf {\bibinfo {volume} {57}},\
  \bibinfo {pages} {4952} (\bibinfo {year} {1998})}\BibitemShut {NoStop}%
\bibitem [{\citenamefont {Dalvit}\ and\ \citenamefont
  {Mazzitelli}(1999)}]{Dalvit2}%
  \BibitemOpen
  \bibfield  {author} {\bibinfo {author} {\bibfnamefont {D.}~\bibnamefont
  {Dalvit}}\ and\ \bibinfo {author} {\bibfnamefont {F.~D.}\ \bibnamefont
  {Mazzitelli}},\ }\bibfield  {title} {\bibinfo {title} {Creation of photons in
  an oscillating cavity with two moving mirrors},\ }\href@noop {} {\bibfield
  {journal} {\bibinfo  {journal} {Phys. Rev. A}\ }\textbf {\bibinfo {volume}
  {59}},\ \bibinfo {pages} {3049} (\bibinfo {year} {1999})}\BibitemShut
  {NoStop}%
\bibitem [{\citenamefont {Michael}\ \emph {et~al.}(2019)\citenamefont
  {Michael}, \citenamefont {Schmiedmayer},\ and\ \citenamefont
  {Demler}}]{Michael}%
  \BibitemOpen
  \bibfield  {author} {\bibinfo {author} {\bibfnamefont {M.~H.}\ \bibnamefont
  {Michael}}, \bibinfo {author} {\bibfnamefont {J.}~\bibnamefont
  {Schmiedmayer}},\ and\ \bibinfo {author} {\bibfnamefont {E.}~\bibnamefont
  {Demler}},\ }\bibfield  {title} {\bibinfo {title} {From the moving piston to
  the dynamical casimir effect: Explorations with shaken condensates},\
  }\href@noop {} {\bibfield  {journal} {\bibinfo  {journal} {Phys. Rev. A}\
  }\textbf {\bibinfo {volume} {99}},\ \bibinfo {pages} {053615} (\bibinfo
  {year} {2019})}\BibitemShut {NoStop}%
\bibitem [{\citenamefont {Engheta}(2021)}]{Engheta1}%
  \BibitemOpen
  \bibfield  {author} {\bibinfo {author} {\bibfnamefont {N.}~\bibnamefont
  {Engheta}},\ }\bibfield  {title} {\bibinfo {title} {Metamaterials with high
  degrees of freedom: space, time, and more},\ }\href@noop {} {\bibfield
  {journal} {\bibinfo  {journal} {Nanophotonics}\ }\textbf {\bibinfo {volume}
  {10}},\ \bibinfo {pages} {639} (\bibinfo {year} {2021})}\BibitemShut
  {NoStop}%
\bibitem [{\citenamefont {Galiffi}\ \emph {et~al.}(2022)\citenamefont
  {Galiffi}, \citenamefont {Tirole}, \citenamefont {Yin}, \citenamefont {Li},
  \citenamefont {Vezzoli}, \citenamefont {Huidobro}, \citenamefont
  {Silveirinha}, \citenamefont {Sapienza}, \citenamefont {Al\`u},\ and\
  \citenamefont {Pendry}}]{galiffi1}%
  \BibitemOpen
  \bibfield  {author} {\bibinfo {author} {\bibfnamefont {E.}~\bibnamefont
  {Galiffi}}, \bibinfo {author} {\bibfnamefont {R.}~\bibnamefont {Tirole}},
  \bibinfo {author} {\bibfnamefont {S.}~\bibnamefont {Yin}}, \bibinfo {author}
  {\bibfnamefont {H.}~\bibnamefont {Li}}, \bibinfo {author} {\bibfnamefont
  {S.}~\bibnamefont {Vezzoli}}, \bibinfo {author} {\bibfnamefont {P.~A.}\
  \bibnamefont {Huidobro}}, \bibinfo {author} {\bibfnamefont {M.~G.}\
  \bibnamefont {Silveirinha}}, \bibinfo {author} {\bibfnamefont
  {R.}~\bibnamefont {Sapienza}}, \bibinfo {author} {\bibfnamefont
  {A.}~\bibnamefont {Al\`u}},\ and\ \bibinfo {author} {\bibfnamefont {J.~B.}\
  \bibnamefont {Pendry}},\ }\bibfield  {title} {\bibinfo {title} {Photonics of
  time-varying media},\ }\href@noop {} {\bibfield  {journal} {\bibinfo
  {journal} {Adv. Photon.}\ }\textbf {\bibinfo {volume} {4(1)}},\ \bibinfo
  {pages} {014002} (\bibinfo {year} {2022})}\BibitemShut {NoStop}%
\bibitem [{\citenamefont {Siegman}(1986)}]{Siegman}%
  \BibitemOpen
  \bibfield  {author} {\bibinfo {author} {\bibfnamefont {A.~E.}\ \bibnamefont
  {Siegman}},\ }\href@noop {} {\emph {\bibinfo {title} {Lasers}}}\ (\bibinfo
  {publisher} {University Science Books, Mill Valley, California},\ \bibinfo
  {year} {1986})\BibitemShut {NoStop}%
\bibitem [{\citenamefont {Costen}\ and\ \citenamefont
  {Adamson}(1965)}]{costen}%
  \BibitemOpen
  \bibfield  {author} {\bibinfo {author} {\bibfnamefont {R.}~\bibnamefont
  {Costen}}\ and\ \bibinfo {author} {\bibfnamefont {D.}~\bibnamefont
  {Adamson}},\ }\bibfield  {title} {\bibinfo {title} {Three-dimensional
  derivation of the electrodynamic jump conditions and momentum-energy laws at
  a moving boundary},\ }\href@noop {} {\bibfield  {journal} {\bibinfo
  {journal} {Proc. IEEE}\ }\textbf {\bibinfo {volume} {53}},\ \bibinfo {pages}
  {1181} (\bibinfo {year} {1965})}\BibitemShut {NoStop}%
\bibitem [{\citenamefont {Garabedian}(1964)}]{Garabedian}%
  \BibitemOpen
  \bibfield  {author} {\bibinfo {author} {\bibfnamefont {P.~R.}\ \bibnamefont
  {Garabedian}},\ }\href@noop {} {\emph {\bibinfo {title} {Partial Differential
  Equations}}}\ (\bibinfo  {publisher} {John Wiley and Sons},\ \bibinfo {year}
  {1964})\BibitemShut {NoStop}%
\bibitem [{\citenamefont {John}(1982)}]{John}%
  \BibitemOpen
  \bibfield  {author} {\bibinfo {author} {\bibfnamefont {F.}~\bibnamefont
  {John}},\ }\href@noop {} {\emph {\bibinfo {title} {Partial Differential
  Equations}}},\ \bibinfo {edition} {4th}\ ed.\ (\bibinfo  {publisher}
  {Springer},\ \bibinfo {year} {1982})\BibitemShut {NoStop}%
\bibitem [{Note1()}]{Note1}%
  \BibitemOpen
  \bibinfo {note} {The equation has opposite signs for the velocity, if the
  characteristics were reflected from the left instead of the right as seen in
  Fig. \ref {fig:fig_3}}\BibitemShut {NoStop}%
\bibitem [{\citenamefont {Cassedy}\ and\ \citenamefont
  {Oliner}(1963)}]{Cassedy1}%
  \BibitemOpen
  \bibfield  {author} {\bibinfo {author} {\bibfnamefont {E.~S.}\ \bibnamefont
  {Cassedy}}\ and\ \bibinfo {author} {\bibfnamefont {A.~A.}\ \bibnamefont
  {Oliner}},\ }\bibfield  {title} {\bibinfo {title} {Dispersion relations in
  time-space periodic media: part i-stable interactions},\ }\href@noop {}
  {\bibfield  {journal} {\bibinfo  {journal} {Proc. IEEE}\ }\textbf {\bibinfo
  {volume} {51}},\ \bibinfo {pages} {1342} (\bibinfo {year}
  {1963})}\BibitemShut {NoStop}%
\bibitem [{\citenamefont {Cassedy}(1967)}]{Cassedy2}%
  \BibitemOpen
  \bibfield  {author} {\bibinfo {author} {\bibfnamefont {E.~S.}\ \bibnamefont
  {Cassedy}},\ }\bibfield  {title} {\bibinfo {title} {Dispersion relations in
  time-space periodic media: part ii-unstable interactions},\ }\href@noop {}
  {\bibfield  {journal} {\bibinfo  {journal} {Proc. IEEE}\ }\textbf {\bibinfo
  {volume} {55}},\ \bibinfo {pages} {1154} (\bibinfo {year}
  {1967})}\BibitemShut {NoStop}%
\bibitem [{\citenamefont {Koutserimpas}(2022)}]{koutserimpas1}%
  \BibitemOpen
  \bibfield  {author} {\bibinfo {author} {\bibfnamefont {T.~T.}\ \bibnamefont
  {Koutserimpas}},\ }\bibfield  {title} {\bibinfo {title} {Parametric
  amplification interactions in time-periodic media: coupled waves theory},\
  }\href@noop {} {\bibfield  {journal} {\bibinfo  {journal} {J. Opt. Soc. Am.
  B}\ }\textbf {\bibinfo {volume} {39}},\ \bibinfo {pages} {481} (\bibinfo
  {year} {2022})}\BibitemShut {NoStop}%
\bibitem [{\citenamefont {Pendry}\ \emph {et~al.}(2021)\citenamefont {Pendry},
  \citenamefont {Galiffi},\ and\ \citenamefont {Huidobro}}]{pendry1}%
  \BibitemOpen
  \bibfield  {author} {\bibinfo {author} {\bibfnamefont {J.~B.}\ \bibnamefont
  {Pendry}}, \bibinfo {author} {\bibfnamefont {E.}~\bibnamefont {Galiffi}},\
  and\ \bibinfo {author} {\bibfnamefont {P.~A.}\ \bibnamefont {Huidobro}},\
  }\bibfield  {title} {\bibinfo {title} {Gain in time-dependent media - a new
  mechanism},\ }\href@noop {} {\bibfield  {journal} {\bibinfo  {journal} {J.
  Opt. Soc. Am. B}\ }\textbf {\bibinfo {volume} {38}},\ \bibinfo {pages} {3360}
  (\bibinfo {year} {2021})}\BibitemShut {NoStop}%
\bibitem [{\citenamefont {Golzalez}(1997)}]{Gonzalez}%
  \BibitemOpen
  \bibfield  {author} {\bibinfo {author} {\bibfnamefont {N.}~\bibnamefont
  {Golzalez}},\ }\emph {\bibinfo {title} {L'\'equation des ondes dans un domain
  d\'ependant du temps}},\ \href@noop {} {\bibinfo {type} {{Ph.D.} thesis}},\
  \bibinfo  {school} {University of Toulon, Czech Technical University of
  Prague} (\bibinfo {year} {1997})\BibitemShut {NoStop}%
\bibitem [{\citenamefont {Smith}(1967)}]{smith1}%
  \BibitemOpen
  \bibfield  {author} {\bibinfo {author} {\bibfnamefont {P.~W.}\ \bibnamefont
  {Smith}},\ }\bibfield  {title} {\bibinfo {title} {Phase locking of laser
  modes by continuous cavity length variation},\ }\href@noop {} {\bibfield
  {journal} {\bibinfo  {journal} {Appl. Phys. Lett.}\ }\textbf {\bibinfo
  {volume} {10}},\ \bibinfo {pages} {51} (\bibinfo {year} {1967})}\BibitemShut
  {NoStop}%
\bibitem [{\citenamefont {de~Sterke}\ and\ \citenamefont
  {Steel}(1995)}]{sterke1}%
  \BibitemOpen
  \bibfield  {author} {\bibinfo {author} {\bibfnamefont {C.~M.}\ \bibnamefont
  {de~Sterke}}\ and\ \bibinfo {author} {\bibfnamefont {M.~J.}\ \bibnamefont
  {Steel}},\ }\bibfield  {title} {\bibinfo {title} {Simple model for pulse
  formation in lasers with a frequency-shifting element and nonlinearity},\
  }\href@noop {} {\bibfield  {journal} {\bibinfo  {journal} {Opt. Commun.}\
  }\textbf {\bibinfo {volume} {117}},\ \bibinfo {pages} {469} (\bibinfo {year}
  {1995})}\BibitemShut {NoStop}%
\bibitem [{\citenamefont {Taflove}\ and\ \citenamefont
  {Hagness}(2005)}]{Taflove}%
  \BibitemOpen
  \bibfield  {author} {\bibinfo {author} {\bibfnamefont {A.}~\bibnamefont
  {Taflove}}\ and\ \bibinfo {author} {\bibfnamefont {S.~C.}\ \bibnamefont
  {Hagness}},\ }\href@noop {} {\emph {\bibinfo {title} {Computational
  Electrodynamics: The Finite-Difference Time-Domain Method}}},\ \bibinfo
  {edition} {3rd}\ ed.\ (\bibinfo  {publisher} {Artech House, Inc.},\ \bibinfo
  {year} {2005})\BibitemShut {NoStop}%
\bibitem [{\citenamefont {Harfoush}\ \emph {et~al.}(1989)\citenamefont
  {Harfoush}, \citenamefont {Taflove},\ and\ \citenamefont
  {Kriegsmann}}]{harfoush}%
  \BibitemOpen
  \bibfield  {author} {\bibinfo {author} {\bibfnamefont {F.}~\bibnamefont
  {Harfoush}}, \bibinfo {author} {\bibfnamefont {A.}~\bibnamefont {Taflove}},\
  and\ \bibinfo {author} {\bibfnamefont {G.~A.}\ \bibnamefont {Kriegsmann}},\
  }\bibfield  {title} {\bibinfo {title} {A numerical technique for analyzing
  electromagnetic wave scattering from moving surfaces in one or two
  dimensions},\ }\href@noop {} {\bibfield  {journal} {\bibinfo  {journal} {IEEE
  Trans. Antennas Propag.}\ }\textbf {\bibinfo {volume} {37}},\ \bibinfo
  {pages} {55} (\bibinfo {year} {1989})}\BibitemShut {NoStop}%
\bibitem [{Note2()}]{Note2}%
  \BibitemOpen
  \bibinfo {note} {Supplementary material with two videos of the electric field
  oscillations for: (a) $T=1$ and (b) $T=0.5$.}\BibitemShut {Stop}%
\bibitem [{\citenamefont {Noronen}\ \emph {et~al.}(2016)\citenamefont
  {Noronen}, \citenamefont {Okhotnikov},\ and\ \citenamefont
  {Gumenyuk}}]{Noronen1}%
  \BibitemOpen
  \bibfield  {author} {\bibinfo {author} {\bibfnamefont {T.}~\bibnamefont
  {Noronen}}, \bibinfo {author} {\bibfnamefont {O.}~\bibnamefont
  {Okhotnikov}},\ and\ \bibinfo {author} {\bibfnamefont {R.}~\bibnamefont
  {Gumenyuk}},\ }\bibfield  {title} {\bibinfo {title} {Electronically tunable
  thulium-holmium mode-locked fiber laser for the 1700-1800 nm wavelength
  band},\ }\href@noop {} {\bibfield  {journal} {\bibinfo  {journal} {Opt.
  Express}\ }\textbf {\bibinfo {volume} {24}},\ \bibinfo {pages} {14703}
  (\bibinfo {year} {2016})}\BibitemShut {NoStop}%
\bibitem [{\citenamefont {Woodward}\ \emph {et~al.}(2018)\citenamefont
  {Woodward}, \citenamefont {Majewski},\ and\ \citenamefont
  {Jackson}}]{Woodward1}%
  \BibitemOpen
  \bibfield  {author} {\bibinfo {author} {\bibfnamefont {R.~I.}\ \bibnamefont
  {Woodward}}, \bibinfo {author} {\bibfnamefont {M.~R.}\ \bibnamefont
  {Majewski}},\ and\ \bibinfo {author} {\bibfnamefont {S.~D.}\ \bibnamefont
  {Jackson}},\ }\bibfield  {title} {\bibinfo {title} {Mode-locked dysprosium
  fiber laser: Picosecond pulse generation from 2.97 to 3.30 micro meters},\
  }\href@noop {} {\bibfield  {journal} {\bibinfo  {journal} {APL Photonics}\
  }\textbf {\bibinfo {volume} {3}},\ \bibinfo {pages} {116106} (\bibinfo {year}
  {2018})}\BibitemShut {NoStop}%
\bibitem [{\citenamefont {Majewski}\ \emph {et~al.}(2019)\citenamefont
  {Majewski}, \citenamefont {Woodward},\ and\ \citenamefont
  {Jackson}}]{Majewski1}%
  \BibitemOpen
  \bibfield  {author} {\bibinfo {author} {\bibfnamefont {M.~R.}\ \bibnamefont
  {Majewski}}, \bibinfo {author} {\bibfnamefont {R.~I.}\ \bibnamefont
  {Woodward}},\ and\ \bibinfo {author} {\bibfnamefont {S.~D.}\ \bibnamefont
  {Jackson}},\ }\bibfield  {title} {\bibinfo {title} {Ultrafast mid-infrared
  fiber laser mode-locked using frequency-shifted feedback},\ }\href@noop {}
  {\bibfield  {journal} {\bibinfo  {journal} {Opt. Lett.}\ }\textbf {\bibinfo
  {volume} {44}},\ \bibinfo {pages} {1698} (\bibinfo {year}
  {2019})}\BibitemShut {NoStop}%
\bibitem [{\citenamefont {Kuizenga}\ and\ \citenamefont
  {Siegman}(1970{\natexlab{a}})}]{Kuizenga1}%
  \BibitemOpen
  \bibfield  {author} {\bibinfo {author} {\bibfnamefont {D.}~\bibnamefont
  {Kuizenga}}\ and\ \bibinfo {author} {\bibfnamefont {A.}~\bibnamefont
  {Siegman}},\ }\bibfield  {title} {\bibinfo {title} {Fm and am mode locking of
  the homogeneous laser - part i: Theory},\ }\href@noop {} {\bibfield
  {journal} {\bibinfo  {journal} {IEEE J. Quantum Electron.}\ }\textbf
  {\bibinfo {volume} {6}},\ \bibinfo {pages} {694} (\bibinfo {year}
  {1970}{\natexlab{a}})}\BibitemShut {NoStop}%
\bibitem [{\citenamefont {Kuizenga}\ and\ \citenamefont
  {Siegman}(1970{\natexlab{b}})}]{Kuizenga2}%
  \BibitemOpen
  \bibfield  {author} {\bibinfo {author} {\bibfnamefont {D.}~\bibnamefont
  {Kuizenga}}\ and\ \bibinfo {author} {\bibfnamefont {A.}~\bibnamefont
  {Siegman}},\ }\bibfield  {title} {\bibinfo {title} {Fm and am mode locking of
  the homogeneous laser - part ii: Experimental results in a nd:yag laser with
  internal fm modulation},\ }\href@noop {} {\bibfield  {journal} {\bibinfo
  {journal} {IEEE J. Quantum Electron.}\ }\textbf {\bibinfo {volume} {6}},\
  \bibinfo {pages} {709} (\bibinfo {year} {1970}{\natexlab{b}})}\BibitemShut
  {NoStop}%
\bibitem [{\citenamefont {Usechak}\ \emph {et~al.}(2011)\citenamefont
  {Usechak}, \citenamefont {Grupen}, \citenamefont {Naderi}, \citenamefont
  {Li}, \citenamefont {Lester},\ and\ \citenamefont {Kovanis}}]{kovanis}%
  \BibitemOpen
  \bibfield  {author} {\bibinfo {author} {\bibfnamefont {N.~G.}\ \bibnamefont
  {Usechak}}, \bibinfo {author} {\bibfnamefont {M.}~\bibnamefont {Grupen}},
  \bibinfo {author} {\bibfnamefont {N.}~\bibnamefont {Naderi}}, \bibinfo
  {author} {\bibfnamefont {Y.}~\bibnamefont {Li}}, \bibinfo {author}
  {\bibfnamefont {L.~F.}\ \bibnamefont {Lester}},\ and\ \bibinfo {author}
  {\bibfnamefont {V.}~\bibnamefont {Kovanis}},\ }\bibfield  {title} {\bibinfo
  {title} {Modulation effects in multi-section semiconductor lasers},\
  }\href@noop {} {\bibfield  {journal} {\bibinfo  {journal} {Proceedings of
  SPIE, Physics and Simulation of Optoelectronic Devices XIX}\ }\textbf
  {\bibinfo {volume} {7933}},\ \bibinfo {pages} {79331I} (\bibinfo {year}
  {2011})}\BibitemShut {NoStop}%
\bibitem [{\citenamefont {Grosso}\ and\ \citenamefont
  {Yellin}(1977)}]{grosso1}%
  \BibitemOpen
  \bibfield  {author} {\bibinfo {author} {\bibfnamefont {R.~P.}\ \bibnamefont
  {Grosso}}\ and\ \bibinfo {author} {\bibfnamefont {M.}~\bibnamefont
  {Yellin}},\ }\bibfield  {title} {\bibinfo {title} {The membrane mirror as an
  adaptive optical element},\ }\href@noop {} {\bibfield  {journal} {\bibinfo
  {journal} {J. Opt. Soc. Am.}\ }\textbf {\bibinfo {volume} {67}},\ \bibinfo
  {pages} {399} (\bibinfo {year} {1977})}\BibitemShut {NoStop}%
\bibitem [{\citenamefont {Al-Mashaal}\ \emph {et~al.}(2018)\citenamefont
  {Al-Mashaal}, \citenamefont {Wood}, \citenamefont {Torin}, \citenamefont
  {Mastropaolo}, \citenamefont {Newton},\ and\ \citenamefont {Cheung}}]{Al1}%
  \BibitemOpen
  \bibfield  {author} {\bibinfo {author} {\bibfnamefont {A.~K.}\ \bibnamefont
  {Al-Mashaal}}, \bibinfo {author} {\bibfnamefont {G.~S.}\ \bibnamefont
  {Wood}}, \bibinfo {author} {\bibfnamefont {A.}~\bibnamefont {Torin}},
  \bibinfo {author} {\bibfnamefont {E.}~\bibnamefont {Mastropaolo}}, \bibinfo
  {author} {\bibfnamefont {M.~J.}\ \bibnamefont {Newton}},\ and\ \bibinfo
  {author} {\bibfnamefont {R.}~\bibnamefont {Cheung}},\ }\bibfield  {title}
  {\bibinfo {title} {Tunable graphene-polymer resonators for audio frequency
  sensing applications},\ }\href@noop {} {\bibfield  {journal} {\bibinfo
  {journal} {IEEE Sens. J.}\ }\textbf {\bibinfo {volume} {19}},\ \bibinfo
  {pages} {465} (\bibinfo {year} {2018})}\BibitemShut {NoStop}%
\bibitem [{\citenamefont {Groeblacher}\ \emph {et~al.}(2009)\citenamefont
  {Groeblacher}, \citenamefont {Hammerer}, \citenamefont {Vanner},\ and\
  \citenamefont {Aspelmeyer}}]{Groeblacher}%
  \BibitemOpen
  \bibfield  {author} {\bibinfo {author} {\bibfnamefont {S.}~\bibnamefont
  {Groeblacher}}, \bibinfo {author} {\bibfnamefont {K.}~\bibnamefont
  {Hammerer}}, \bibinfo {author} {\bibfnamefont {M.~R.}\ \bibnamefont
  {Vanner}},\ and\ \bibinfo {author} {\bibfnamefont {M.}~\bibnamefont
  {Aspelmeyer}},\ }\bibfield  {title} {\bibinfo {title} {Observation of strong
  coupling between a micromechanical resonator and an optical cavity field},\
  }\href@noop {} {\bibfield  {journal} {\bibinfo  {journal} {Nature}\ }\textbf
  {\bibinfo {volume} {460}},\ \bibinfo {pages} {724} (\bibinfo {year}
  {2009})}\BibitemShut {NoStop}%
\bibitem [{\citenamefont {Anetsberger}\ \emph {et~al.}(2009)\citenamefont
  {Anetsberger}, \citenamefont {Arcizet}, \citenamefont {Unterreithmeier},
  \citenamefont {Riviere}, \citenamefont {Schliesser}, \citenamefont {Weig},
  \citenamefont {Kotthaus},\ and\ \citenamefont {Kippenberg}}]{Anetsberger}%
  \BibitemOpen
  \bibfield  {author} {\bibinfo {author} {\bibfnamefont {G.}~\bibnamefont
  {Anetsberger}}, \bibinfo {author} {\bibfnamefont {O.}~\bibnamefont
  {Arcizet}}, \bibinfo {author} {\bibfnamefont {Q.~P.}\ \bibnamefont
  {Unterreithmeier}}, \bibinfo {author} {\bibfnamefont {R.}~\bibnamefont
  {Riviere}}, \bibinfo {author} {\bibfnamefont {A.}~\bibnamefont {Schliesser}},
  \bibinfo {author} {\bibfnamefont {E.~M.}\ \bibnamefont {Weig}}, \bibinfo
  {author} {\bibfnamefont {J.~P.}\ \bibnamefont {Kotthaus}},\ and\ \bibinfo
  {author} {\bibfnamefont {T.~J.}\ \bibnamefont {Kippenberg}},\ }\bibfield
  {title} {\bibinfo {title} {Near-field cavity optomechanics with
  nanomechanical oscillators},\ }\href@noop {} {\bibfield  {journal} {\bibinfo
  {journal} {Nature Physics}\ }\textbf {\bibinfo {volume} {5}},\ \bibinfo
  {pages} {909} (\bibinfo {year} {2009})}\BibitemShut {NoStop}%
\bibitem [{\citenamefont {Friend}\ and\ \citenamefont {Yeo}(2011)}]{Friend}%
  \BibitemOpen
  \bibfield  {author} {\bibinfo {author} {\bibfnamefont {J.}~\bibnamefont
  {Friend}}\ and\ \bibinfo {author} {\bibfnamefont {L.~Y.}\ \bibnamefont
  {Yeo}},\ }\bibfield  {title} {\bibinfo {title} {Microscale acoustofluidics:
  Microfluidics driven via acoustics and ultrasonics},\ }\href@noop {}
  {\bibfield  {journal} {\bibinfo  {journal} {Rev. Mod. Phys.}\ }\textbf
  {\bibinfo {volume} {83}},\ \bibinfo {pages} {647} (\bibinfo {year}
  {2011})}\BibitemShut {NoStop}%
\end{thebibliography}%

\end{document}